\documentstyle[12pt, psfig,epsf]{article}
\begin{document}

\begin{center}
{\Large \bf DIANA, a program for Feynman Diagram Evaluation} \\

\vspace{4mm}
M.~Tentyukov\footnote{On leave of absence from
Joint Institute for Nuclear Research 141980 Dubna, 
Moscow Region, Russian Federation.
E-mail:
tentukov@physik.uni-bielefeld.de },
J.~Fleischer\footnote{E-mail: fleischer@physik.uni-bielefeld.de}\\
Fakult\"at f\"ur Physik, Universit\"at Bielefeld\\
D-33615 Bielefeld, Germany.

\begin{abstract}
A C-program
DIANA (DIagram ANAlyser) for the automatic Feynman diagram 
evaluation is presented. It consists of two parts:
the analyzer of diagrams and 
the interpreter of a special text manipulating language.
This language is used to create a source code 
for analytical or numerical evaluations and to keep the control of
the process in general.
\end{abstract}

\end{center}

   Recent high precision experiments require, on the side of the theory,
high-precision calculations resulting in the evaluation of higher
loop dia\-grams in the Standard Model (SM).
For specific processes thousands of multiloop Feynman
dia\-grams do contribute. For example, in the calculation of the
anomalous magnetic
moment of the muon $\frac{1}{2}(g-2)_\mu$ there are 1832 diagrams in
two loops. Of course, the contribution of most of these
diagrams is very small. But sometimes it is not so easy to distinguish
between important and unimportant diagrams. On the other hand, we
often need to take into account all diagrams, to verify gauge
independence, or cancellation of divergences. 
It turns out impossible to perform
these calculations by hand. This makes the request for automation a
high-priority task. 

Several different packages
have been developed with different areas of applicability ( for a review 
see \cite{Steinh1}).
For example, there are several well-known Mathema\-tica packages:
FeynArts / FeynCalc \cite{FeynmArts} are convenient
for various aspects of the calculation of radiative corrections in the
SM, TARCER \cite{tarcer} - a program for the reduction of two-loop
propagator integrals.
There are also several FORM packages for evaluating multiloop diagrams, like
MINCER \cite{MINCER}, and a package \cite{leo96} 
for the calculation of 3-loop bubble integrals with one non-zero mass.
Other packages for automation are
GRACE \cite{GRACE} and CompHEP \cite{CompHep},
which partially perform full calculations, from the process definition 
to the cross-section values. 

A somewhat different approach is pursued  by
XLOOPS \cite{XLoops}. 
A graphical user interface
makes XLOOPS an `easy-to-handle' program package.
Recent development of XLOOPS is aimed at the  evaluation of all
diagrams occurring in given processes, but at present
it is  mainly aimed 
to the evaluation of single diagrams. 

To deal with thousands of diagrams, it is necessary to use 
special techniques like databases and special controlling programs.
In \cite{Vermaseren} for evalua\-ting more than 11000 diagrams
the special database-like program MINOS was developed.
It calls the relevant FORM programs, waits until they fi\-nished, picks 
up their results and repeats the process without any human interference.

It seems  impossible to develop an  universal package, which 
will be efficient for all tasks.
Various ways will be of different
efficiency, have different domains of applicability, and
should eventually allow for completely independent checks of
the final results.
This point of view motivated us to seek our own
way of automatic evaluation of Feynman diagrams.

To understand the idea, let us look at 
the new package  GEFICOM \cite{GEFICOM}.

This package was developed for computation 
of higher order processes involving a large number of diagrams.
It is based on cooperative use of several software tools such as 
Mathematica, FORM, FORTRAN, etc.

The obvious advantage of such an approach is that all partial tasks can
be solved by means of the most convenient software tool.

But many kinds of the system-depending software lead to problems:

1)~Problems with portability: such a system requires use of many
operational system resources and strongly depends on them.
E.g, the most popular family OS MS Windows has not enough abilities.
2)~Problems with the installation process for non-specialists.
3)~Usually, such a system is elaborated ``ad hoc'', and it is very
difficult to change its general structure.

Our aim is to create some  universal
software tool for piloting the process of generating the source
code  in multi-loop order
for analytical or numerical evaluations  and to keep the control of
the process  in general. Based on this instrument, we can attempt to build 
a complete package performing the computation of any given process
in the framework of any concrete model. 

It is worth noting that such approach is widely used nowadays.
For example, the well-known package FormCalc \cite{FormCalc}
uses Mathematica as a front-end for FORM. Combining the advantages of
Mathematica and FORM, this package is well suited for the big one-loop
level problems.

But, Mathematica is too big and slow to guarantee controlling
facility for the calculation of a huge process. From this point of
view, it is overloaded by many unnecessary built-in functions; it
was elaborated for solving a  bit other problems. It is not so
efficient when it is used as a shell language.

There are many  software tools elaborated  specially for these
tasks. Usually, the ``make'' utility is used combining with some shell
language, like the PERL interpreter. Here, again, problems specific for
heterogeneous systems appear.
And, many languages in use make difficulties for non-specialists to
deal with such a system. 

We need the ``scripting'' language suitable for both controlling the
various software and generating the source code for this software. All
existing languages either are not powerful enough or too complicated for
non-specialists. 

For the project called  DIANA (DIagram ANAlyser) \cite{Diana} for the
evalua\-tion of Feynman diagrams we have elaborated a special 
text manipulating language (TM).

The TM language is a  TeX-like language for creating
source code, organizing the interactive dialog and so on. It is very
simple, but, nevertheless, it is very configurable and extensible.

Similar to the TeX language, all lines without special escape - 
characters (``$\backslash$'') are simply typed to the output file. 
So, to type
``Hello, world!'' to the terminal we may write down the following
program:
\small\begin{verbatim}
Hello, world!
\end{verbatim}\normalsize

Each word,
the first character of which is the escape character, will 
be considered as a command. There are many built-in commands returning 
the information about diagrams. For example, if the user writes in the input
file 
\small\begin{verbatim}
\mass(2),
\end{verbatim}\normalsize
in the output file
the value of the particle mass on line number 2  appears.

For generating Feynman diagrams we use QGRAF \cite{QGRAF}. 
DIANA reads QGRAF output. For each diagram it performs
the TM-program, producing input for further evaluation of the diagram.
Thus the program:

Reads QGRAF output and for each diagram it:
1)~Determines the topology,
looking for it in the table of
all known topologies and distributes momenta.
If we do not yet know all needed topologies,
we may use the program
to determine missing topologies that occur in the process.
2)~Creates an internal representation of the diagram 
in terms of vertices and propagators.
3)~Executes the TM-program 
to insert explicit expressions for the vertices, propagators etc.
The TM-program produces input text for FORM ( or some other language),
 and executes the latter (optionally).

Using the TM language, advanced users can develop further extensions,
e.g. including FORTRAN, to create
a postscript file for the  picture of the current diagram, etc.

The program operates as follows:
first of all, it reads its configuration 
file, which may be produced manually or by DIANA as well. This file contains:
1)~The information about various settings (file names, numbers
of external particles, definition of key words, etc.)
2)~Momenta distribution for each topology.
3)~Description of the model (i.e., all particles, propagators and vertices).
4)~TM-program.

The TM-program is part of the configuration file. It starts with the
directive 
\small\begin{verbatim}
\begin translate
\end{verbatim}\normalsize

Then the program starts to read QGRAF output.
For each diagram it determines the topology,
assigns indices and creates the textual representation
of the diagram corresponding to the Feynman integrand. All defined data
(masses of particles, momenta on each lines, etc.) are stored
in internal tables, and may be called by TM-program operators.
At this point DIANA performs the TM-program.
After that it starts to work with the next diagram.

When all diagrams are processed, the program may perform the
TM-program a last time (optionally).
This may be used to do some final operations like summing up the
results.

Use of the TM language makes DIANA very flexible. It is easy to work out
various algorithms of diagram evaluation by specifying settings 
in the configuration file or even by a TM program. 

Let us suppose that we use FORM as formulae manipulating language.

The user types his FORM program and has the possibility to insert 
in the same line TM-operators as well.
For example, the typical part of a TM program looks like follows:
\small\begin{verbatim}
\program
\setout(d\currentdiagramnumber().frm)
#define dia "\currentdiagramnumber()"
#define TYPE "\type()"
#define COLOR "\color()"
#define LINES "\numberofinternallines()"
\masses()
#include def.h
l  R=\integrand();
#call feynmanrules{}
#call projection{}
#call reducing{'TYPE'}
#call table{'TYPE'}
#call colorfactor{'COLOR'}
.sort
g dia'dia' = R;
drop R;
.store
save dia'dia'.sto;
.end
\setout(null)
\system(\(form -l )d\currentdiagramnumber().frm)
\end{verbatim}\normalsize

This TM-program will generate the FORM input for each diagram.
For example, the corresponding part of 
the FORM program generated for diagram number 15 will be placed into the
file ``d15.frm'' and looks like follows (for the detailed explanation of this
example see~\cite{Diana},  Appendix A):
\small\begin{verbatim}
#define dia "15"
#define TYPE "4"
#define COLOR "3"
#define LINES "4"
#define m1 "mmH"
#define m2 "mmW"
#define m3 "mmW"
#define m4 "mmH"
#include def.h
l  R=
       1*V(1,mu1,mu,2)*(-i_)*em^2/2/s*V(2,0)*(-i_)*1/4*em^2/s^2*mmH/mmW*
       V(3,mu2,+q4-(+q3),1)*(-i_)*em/2/s*SS(1,0)*i_*VV(2,mu1,mu2,+q2,2)*i_*
       SS(3,2)*i_*SS(4,0)*i_;
#call feynmanrules{}
#call projection{}
#call reducing{'TYPE'}
#call table{'TYPE'}
#call colorfactor{'COLOR'}
.sort
drop R;
g dia'dia' = R;
.store
save dia'dia'.sto;
.end
\end{verbatim}\normalsize

Then this program will be executed by FORM 
by means of the operator
\small\begin{verbatim}
\system(\(form -l )d\currentdiagramnumber().frm)
\end{verbatim}\normalsize
In this case the operator performs the command
\small\begin{verbatim}
form -l d15.frm
\end{verbatim}\normalsize{}

There is a possibility to use DIANA to perform the TM-program only,
without reading QGRAF output.
If one specifies in the configuration 
file \small\verb|only interpret|\normalsize{}
then DIANA will not try to read QGRAF output, but immediately
enters the TM - program.

DIANA contains a powerful preprocessor. The user can create macros to hide 
complicated constructions.
Similar as LaTeX provides the possibility for non-specialists to
typeset high-quality texts using the TeX language, 
these macros permit DIANA to work at very high level. The user
can specify the model and the process, and DIANA will generate all necessary 
files \cite{Diana}. 

At present DIANA is available only upon request from the authors.

\section*{Acknowledgements}
This work has been supported by the DFG project FL241/4-1 and in part by 
RFBR \#98-02-16923.

\end{document}